\documentclass[a4paper]{article}
\usepackage[natbibapa]{apacite}
\usepackage[margin=3cm]{geometry}
\usepackage{graphicx}
\usepackage[hidelinks]{hyperref}
\title{Automatic de-identification of Data Download Packages}
\date{}
\author{Laura Boeschoten, Roos Voorvaart, Casper Kaandorp, \\
Ruben van den Goorbergh \& Martine de Vos}

\begin{document}

\maketitle

\begin{abstract}
The General Data Protection Regulation (GDPR) grants all natural persons the right of access to their personal data if this is being processed by data controllers. The data controllers are obliged to share the data in an electronic format and often provide the data in a so called Data Download Package (DDP). These DDPs contain all data collected by public and private entities during  the course of citizens' digital life and form a treasure trove for social scientists. However, the data can be deeply private. To protect the privacy of research participants while using their DDPs for scientific research, we developed de-identification software that is able to handle typical characteristics of DDPs such as regularly changing file structures, visual and textual content, different file formats, different file structures and accounting for usernames. We investigate the performance of the software and illustrate how the software can be tailored towards specific DDP structures.
\end{abstract}




\section{Introduction}\label{s1}

The General Data Protection Regulation (GDPR) grants all natural persons the right of access to their personal data if  this is being processed by data controllers, such as tech companies, governments and mobile phone providers \citep{regulation2016regulation}. Data controllers are obliged to provide a copy of this personal data in a machine readable format and most large data controllers currently comply with this by providing users with the option to retrieve an electronic ``Data Download Package" (DDP). These DDPs contain all data collected by public and private entities during the course of citizens' digital life and form a new treasure trove for social scientists \citep{king2011ensuring,boeschoten2020digital}. However, depending on which data controller is used, the data collected through DDPs can be deeply private and potentially sensitive. Therefore, collecting DDPs for scientific research raises serious privacy concerns and it would not be in line with the principles listed in the GDPR if appropriate measures to protect the privacy of research participants donating their DDPs would not be taken.

To protect privacy of research participants while using DDPs for scientific research, different types of security measures should be taken such as using shielded (cloud)environments to store the data and using privacy-preserving algorithms when analyzing the data. One key issue here is that the privacy of the participants should be preserved while their data is investigated by researchers and that, although appropriate security measures are taken to prevent this, in case of a data breach, it should not be possible to identify research participants. Because of these reasons, a thorough de-identification procedure is imperative. Many different types of software are already available for this, such as DEDUCE \citep{menger2018deduce} and `de-identify Twitter' \citep{Coppersmith2017Twitter}. However, existing methods are not able to handle the highly complex and unstructured nature of DDPs. A particular characteristic of DDPs, that a de-identification procedure should consider, is the fact that the primary identifier of a natural person can be different for different DDPs and is often a username. Furthermore, some DDPs store private interactions of research participants with their contacts, which should be de-identified as well. At last, in case of personal data protected by the GDPR, `machine readable' unfortunately does not mean equally structured nor easy to parse. Due to this great variety in content and structure, a new method for de-identification of DDPs is essential.

In this research project we developed an automatic de-identification approach that can deal with the variety in DDPs. In the development we focused on DDPs from Instagram but we believe that our approach forms the basis of the de-identification of most DDPs and can easily be extended in order to de-identify DDPs from other companies. 

Our contributions are the following:
\begin{itemize}
    \item We give insight in the structure and content of Instagram DDPs.
    \item We have developed a de-identification algorithm and provide it open source.
    \item We have created an evaluation data set and provide it open source.
    \item We prove that our algorithm is able to find and de-identify a substantive amount of personal data within DDPs. 
    \item We provide the validation algorithm and ground truth used open source.
\end{itemize}

In the Background section we describe in more detail the structure of DDPs and we discuss how privacy of research subjects can be preserved when their DDPs are used for scientific research. In the Methods section we describe our de-identification strategy and how we deal with variety in Instagram DDPs. In addition, this section contains a description of the algorithm that we developed. In the Evaluation section we describe the creation of the evaluation data set. In the Results section we describe the outcomes of this evaluation procedure.

\section{Background}\label{s2}
The aim of the software introduced in this paper is to enable researchers to use DDPs for scientific research while preserving the privacy of participants. In this section, we explain in more detail the specific type of data that can be found in DDPs, define our aims in terms of data protection in more detail and discuss relevant existing literature and software. 

\subsection{Data Download Packages}\label{s2s1}
Most large data controllers currently comply with the right of data access by providing users with the option to retrieve an electronic ``Data Download Package" (DDP). This DDP typically comes as a .zip-file containing .json, .html, .csv, .txt, .JPEG and/or .MP4 files in which all the digital traces left behind by the data subject with respect to the data controller are stored. The structure and content of a DDP varies per data controller, and even within data controllers there are differences among data subjects. Data subjects may use different features provided by the data controller and this is reflected by their DDP, for example, if a data subject does not share photos on Facebook, there will be no data folder with .JPEG files in the corresponding DDP. 

One particular characteristic of DDPs is that their content and structure is often subject to change. For example, if a data subject downloads the DDP at a data controller, and repeats this a month later, differences may be found in the structure of the DDP. This can have several causes. The most straightforward cause is that the data subject generated additional data throughout this month. However, other important factors also play a role. First, data controllers can develop new features by which new types of data regarding the data subject are collected. Second, other features are phased out. Third, some data (for example search history) is only saved for a limited amount of time and is destroyed by the data controller after that period. In that case, it will also not be present in the DDP anymore. At last, the GDPR is still relatively new and data controllers continue to optimize the processes used to transfer the relevant data to its subjects, leading to changes in the structure of DDPs. 

\subsection{Instagram DDPs}\label{s2s2}
As the software in this research project was initially developed to de-identify Instagram DDPs, the structure of these DDPs has been thoroughly investigated. Instagram DDPs come as one or multiple zipfiles (depending on the amount of data available on the data subject). The .zip-file contains a number of folders in which all the visual content is stored, namely ``photos", ``videos", ``profile" and ``stories". The different folders refer to the different Instagram features used by the data subject to generate the visual content. For example, in the folder ``profile", a subject's profile picture can be found, while in the folder ``stories", visual content can be found generated using the ``stories" feature in Instagram, a form of ephemeral sharing. All textual information is collected in a number of .json files. Some of these files have a simple list structure. For example the file ``likes.json" lists all the `likes' given by the subject, supplemented with a timestamp and the username of the Instagram account to which the `like' was given. Files such as `connections.json', `searches.json' and `seen\_content.json' have similar structures. Other files, such as `profile.json' are typically shorter in size but have a more complex structure, as they typically contain different auxiliary characteristics. Other files with such a structure are for example `account\_history.json', `devices.json' and `settings.json'. However, a substantial number of files contains data that is less structured. Examples of such files are `comments.json', `media.json', `messages.json' and `stories\_activities.json'. Furthermore, data subjects at Instagram are not necessarily natural persons. Data subjects at Instagram can be identified by a single and unique Username. Typically, natural persons have individual accounts with an accompanying username, but other institutions, such as for example retail shops or bands can also have an individual account with an accompanying username.

To summarize, software to de-identify Instagram DDPs should be able to handle:
\begin{itemize}
    \item An ever changing file structure
    \item Both visual and  textual content
    \item Different file formats
    \item Files in highly structured and highly unstructured format and different variants in between. 
    \item Natural persons and other users which are identified by their unique username.
\end{itemize}

\subsection{Presevering privacy of research subjects}\label{s2s3}
If DDPs are collected for research purposes, researchers are also considered data controllers and the GDPR applies to them as well \citep[p.95]{van2020general}. Among other things, they are obliged to take technical and organisational security measures aiming to minimise the risk of data abuse \citep[p.112]{van2020general}.

To determine what type of security measures are exactly appropriate in a situation where DDPs are collected for scientific research, the content of the DDPs and the purpose of the research play an important role. DDPs can contain various types of data. It can be structured or unstructured and can come in many different types of formats. Each researcher can be interested in a different aspects of the DDPs, depending on their research question. One researcher might be interested in the frequency of social media use during a Covid-19 lockdown \citep{zhong2021mental}, and uses Instagram DDPs to investigate this. Another researcher might be interested political opinion and electoral success \citep{jungherrAnalyzingPoliticalCommunication2015} \citep{schoenPowerPredictionSocial2013} and uses Twitter DDPs. A third researcher might be interested in personality profiling using Facebook ``likes" \citep{kosinskiPrivateTraitsAttributes2013}.

As can be seen from these examples, some researchers are interested in text, while others are interested in likes or visual content. Consider the situation of a researcher interested in extracting measures of political opinions from text found in DDPs in more detail. Although political opinion is considered a category of sensitive personal data \citep[p.79]{van2020general}, they are allowed to be collected when necessary for scientific research purposes \citep[p.85]{van2020general}. However, as discussed, the researcher collecting this data is obliged to take appropriate security measures such as incorporating data protection measures by design and by default.

Although the sensitive personal data is typically essential for the researcher, this is not necessarily the information from which identification of research subjects can occur. Research subject identification from a DDP in case of a data breach is much more likely to occur due to the direct personal data that can be found within a DDP. However, direct personal data is less likely to be relevant for the research. Therefore, incorporating a step to remove direct personal data from DDPs in the data processing phase when collecting DDPs for research purposes reduces the probability that a research subject is identified in case of a data breach while it will not affect the quality of the data needed to answer the research question. 

\subsection{Related work}\label{s2s4}
To remove direct personal data from DDPs, the software should be able to adhere to the five key characteristics of DDPs introduced in the previous subsection. A first step is to investigate to what extent existing software and literature is able to remove direct personal data from DDPs. A well-known approach is $k-$anonymity \citep{sweeney2002k} which requires that each record in a data-set is similar to at least $k-1$ other records on the potentially identifying variables \citep{el2008protecting}. However, parts of the DDPs are highly unstructured and thereby unique per DDP and reaching $k-$anonymity is therefore not feasible. Much research has focused on the de-identification of electronic health records, for example to enable their use in multi-center research studies \citep{kushida2012strategies}. Scientific open source de-identification tools are available such as DEDUCE \citep{menger2018deduce} as well as commercial tools, such as Amazon Comprehend \citep{simon2018amazon} and CliniDeID \citep{liu2020identifying} \citep{heider2020comparative}. Similar initiatives have taken place to de-identify personal data in other types of data, such as for human resource purposes \citep{vanevaluating}. However, textual content generated from structured data-bases such as for electronic health records or human resources typically have a higher level of structure compared to DDPs and does not handle key identifying information in DDPs, such as usernames or visual content and therefore existing software was not sufficient for our purpose. Alternatively, software has been developed focusing on the removal of usernames, for example for Twitter data \citep{Coppersmith2017Twitter}. Furthermore, many different types of both open source and commercial software are available to identify and blur faces on images and videos, such Microsoft Azure \citep{Microsoft2021Azure}, and Facenet-PyTorch \citep{esler2019}. However, none of the investigated software was able to handle both textual and visual content and both structured and unstructured data within one procedure.

To summarize, a de-identification procedure is required that works appropriately when file structures change rapidly over time, while there are substantive differences in the level of structure within the files, that is able to handle different file formats, that is able to handle both visual and textual content and that recognizes the username as the primary identifier for natural persons, while other types of person identifying information (PII) should also be accounted for, such as first names, phone numbers and e-mail addresses. The developed software aims for such a level of protection that the privacy of the DDP owners (the participants) is always preserved. Importantly, the goal is not to prepare the DDPs for public sharing, however, in the unlikely event of a data breach, the individual research participants should not be directly identifiable. Therefore, the de-identification procedure introduced here should always be supplemented with other security measures such as using a shielded (cloud)environment to store the data and using privacy-preserving algorithms when analyzing the data. 

\section{Method} \label{s3}
In this section we describe the approach and implementation of our de-identification algorithm. The developmental corpus for our algorithm is a small set of DDPs downloaded by the researchers.  Although this data-set was small, we could already see a lot of variety in structure and content providing a useful basis for developing and testing our de-identification approach. All software is written in python and publicly available at \url{https://github.com/UtrechtUniversity/anonymize-ddp}.

\subsection{Approach}
\begin{figure}[t]
\includegraphics[width=\textwidth]{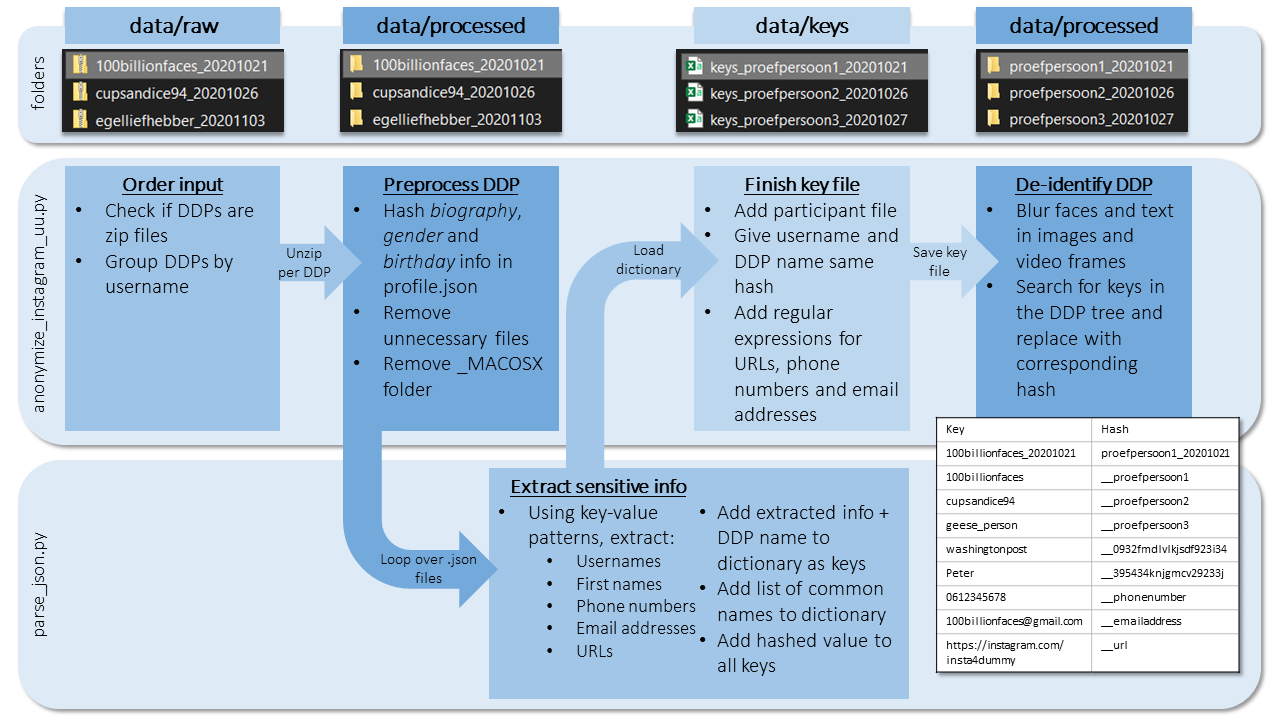}
\caption{The software takes zipped DDP as input. Looping over the text (.json) files, all unique instances of PII are detected in the structured part of the data using pattern- and label recognition. The extracted info, together with the most common Dutch first names and, optionally, the participant file, is added to a key file. All occurrences of the keys in the DDP will be replaced with the corresponding hash. Finally, occurrences of human faces and text in media files are detected and blurred. The software will return a de-identified copy of the DDP in the output folder.}\label{fig:anonymize}
\end{figure}

To de-identify a number of Instagram DDPs, three main steps are undertaken per DDP (see also Figure \ref{fig:anonymize}):
\begin{enumerate}
    \item Preprocess DDP 
    \item De-identify text files:
    \begin{itemize}
        \item Detecting PII in structured text 
        \item Replacing PII with corresponding de-identification codes
    \end{itemize}
    \item De-identify media files by detecting and blurring human faces and text
\end{enumerate}

\subsection{Preprossessing}
The software consists of a wrapper and de-identification algorithms. The wrapper handles the pre-processing of the DDP and contains steps specific for Instagram. It unpacks the DDP and removes all files that are not considered relevant for social science reseach, like ``autofill.json" and ``account history.json". The user's profile ``profile.json" is de-identified separately in this pre-processing phase, as its content and structure deviate from the other text files in the DDP. After the DDP is cleaned, the PII needs to be extracted.

\subsection{De-identify text files}
\subsubsection{Detecting PII in structured text}

\begin{figure}[t]
\includegraphics[scale=0.4]{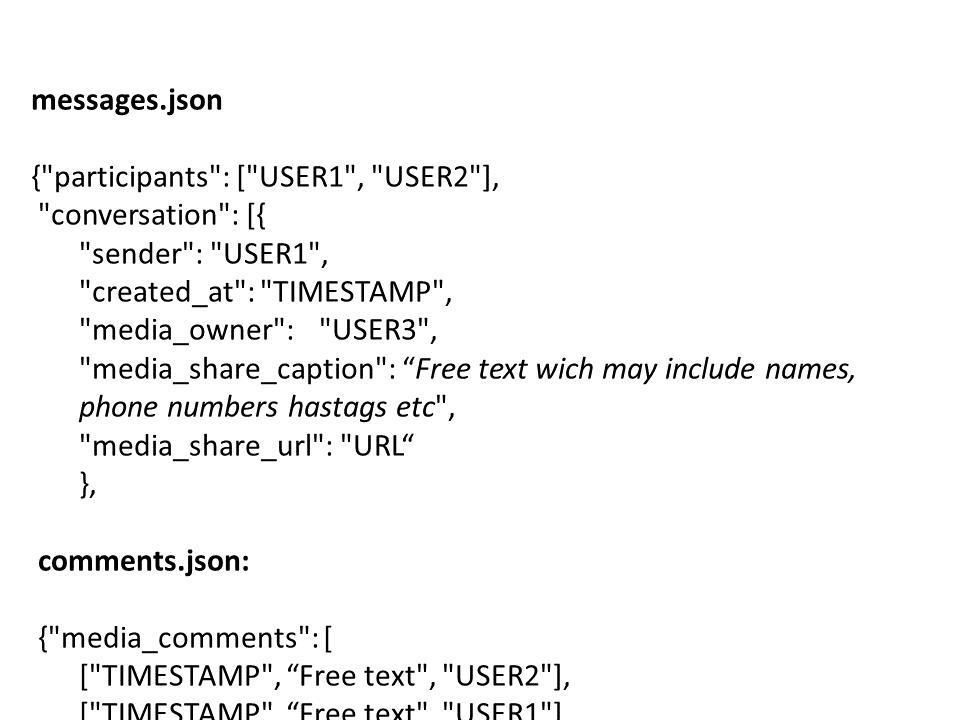}
\caption{Example of key-value structure in .json files with structured and unstructured text.}\label{fig:struct_text}
\end{figure}

All text files in an Instagram DDP contain a nested structure of keys and values (see Figure \ref{fig:struct_text}). To extract PII from these texts, we have determined which key and value combinations and patterns are indicative for PII. 

Per .json file, the algorithm is recursively parsed over the nested structure, each time checking if the specific structure matches (1) a label: username value combination, (2) a username label: timestamp value combination, or (3) a list of length X with at least one timestamp and username value. 

To illustrate the first pattern, each conversation between two or more users stored in the ``messages.json" file is a dictionary, containing multiple sub-dictionaries per sent message. Within this `smallest structure' there is always a label `sender' followed by the username. The algorithm will look for `sender' and other similar standard labels. When the corresponding value matches a username (i.e., a string between 3 and 30 elements without special characters except underscores or points), it will be added to the dictionary.

The second situation can be found in the ``connections.json" file, a dictionary with multiple types of connection labels (e.g., `close\_friends'). Subsequently, each label is made up of another dictionary with all corresponding usernames as labels and timestamps (moment of connection) as values. If the label matches a username and the value a timestamp, the username labels will be saved to the dictionary.

Finally, an example of the (most occurring) third option is the ``comments.json" file. Here you have the various commenting labels (e.g., `media\_comments'), each containing a list of lists. The smallest structure in this file is a list with the time of the comment, the comment, and the username of the owner of the media. After checking if one of the items is a timestamp, the algorithm checks if one of the other items matches a username pattern. If this is the case, the username will be added to the dictionary. 

It should be noted that there is also a fourth way of extracting usernames. Even though most usernames found in Instagram DDPs match the above described patterns, usernames can also be mentioned in free text. In this case, there is no standard pattern to look for. Therefore, using regular expressions, the algorithm will search for tagged people (i.e., `@username') and shared media (i.e., `Shared username’s story') using regular expressions. 

Similar to usernames, the text files are checked for patterns (i.e., `label: PII') and free text indicative of email-addresses and phone numbers. Different from extracting usernames, the regular expressions used to find email-addresses, phone numbers, and URLs are not applied in the `PII-identifying phase', but are explicitly added to the final dictionary. This way, not all occurrences will be added to the dictionary, increasing its size and reducing the efficiency (during the de-identification phase (see below), the algorithm needs to look for each key separately). Instead, by only adding the search patterns to the dictionary, the de-identification process remains efficient and becomes more inclusive. An important side note is that the regular expressions will only look for Instagram URLs. This because most of the URLs in the DDPs represent links to public websites. These cannot be traced to an individual person and they might be valuable for social science research. Therefore, these URLs can be left unchanged.

As (first) names exclusively occur in free text and not in a structured format, it was not possible to systematically extract this type of PII. Therefore, instead of working bottom-up, we applied a top-down approach. After all text files have been checked and the key dictionary is filled, a list of the $10,000$ most common Dutch names is added to this dictionary (which we obtained from the DEDUCE software \citep{menger2018deduce}). Of course, it is also possible to add another list (of another country), making the algorithm applicable in multiple languages. 

\subsubsection{De-identifying PII in text}
After the PII is extracted and added to the dictionary, a PII specific de-identification needs to be added. Usernames and names receive a unique hexadecimal code. Note that the same name will always receive the same code. This way it is still possible to perform a network analysis after anonymization is complete. Additionally, it is also possible to provide the algorithm with a list of (user)names (and/or other information) and specific their corresponding codes yourself. This might be interesting for scientific research in which the (user)names of participants need to be (clearly) distinguishable from other (user)names. In short, (user)names are pseudonymized as they all receive their own specific code and can, therefore, be reverted back if the dictionary is saved. It is up to the user to decide if this dictionary is saved. On the other hand, email-addresses, phone numbers and URLs will anonymized, as they will be hashed with the general `\_\_emailaddress', `\_\_phonenumber', and `\_\_url' codes, respectively. 

For each DDP, the algorithm will look per PII listed in the dictionary for its occurrences, and replace it with the corresponding de-identification code. The replacement extends from file content to file/folder names, resulting in an entirely de-identified DDP. 

\subsection{De-identifying PII in media}
Besides being able to link textual data to specific individuals, individuals may also be identified by their presence in the images or videos in a DDP. In addition, the images or videos can contain text which may include usernames, person names or other sensitive information. We detect faces in visual content using multi-task Cascaded Convolutional Networks \cite{zhang2016facial} in Facenet Pytorch \cite{esler2019} and blur all occurrences using the Python Imaging Library \cite{vankemenade2020}. We detect text using a pre-trained \cite{Yadong2018} EAST text detection model \cite{zhou2017east} and blur all occurrences using the Gaussian blur option provided by OpenCV \cite{opencv2000}.

\section{Evaluation}
\subsection{Data-set}
To evaluate the performance of the software introduced in the Methodology Section, a group of $11$ participants generated Instagram DDPs by actively using a new Instagram account for approximately a week. Here, the participants followed guidelines instructing them to actively generate the type of information that the software aims to de-identify.

The participants were instructed not to share any of their personal information via the Instagram accounts. Instead, participants were instructed to share either fake or publicly available information, such as URLS of news websites, images of celebrities or likes and follows of verified Instagram accounts. As the final data-set does not contain any personal information it is publicly available at \url{http://doi.org/10.5281/zenodo.4472606}.

\begin{table}[ht]
\begin{tabular}{llrrrrrr}
\hline \hline
Visual \\ \hline
&                       & Direct    & Photos    & Profile   & Stories   & Videos    & Total \\ \hline
Files \\
& .JPEG                   & 11        & 525       & 11        &  176      & -         & 723 \\
& .MP4                    & -         & -         & -         & 92        & 15        & 107 \\ \hline
Faces \\
& .JPEG                   &  20       &1046       & -         & 290       & -         & 1,356 \\
& .MP4                    &  -        & -         & -         & 163       & 36        & 199 \\ \hline
Usernames \\
& .JPEG                   & -         & 49        & -         & 255       & -         & 304 \\
& .MP4                    & -         & -         & -         & 105       & 21        & 126 \\ \hline \hline
Textual\\ \hline
&                         DDP\_id  & E-mail & Name   & Phone   & URL   & Username   & Total \\ \hline
comments.json             & -        & 28     &	105   &	29      & 1	    & 261        & 424  \\
connections.json          & -        & -      & -     & -       & -     & 1,222      & 1,222\\
likes.json                & -        & -      & -     & -       & -     & 883        & 883  \\
media.json                & -        & 28	  & 54	  & 9		& -     & 43         & 134  \\
messages.json             & 294	     & 152	  & 421	  & 139	    & 267	& 2,659      & 3,932\\
profile.json              & 18	     & 10	  &	-     & -       & 10	& 1          & 39   \\
saved.json                & -        & -      & -     & -       & -     & 6          & 6    \\
searches.json             & -        & -      & -     & -       & -     & 314        & 314  \\
seen\_content.json        & -        & -      & -     & -       & -     & 3,143      & 3,143\\
shopping.json             & -        & -      & -     & -       & -     & 1          & 1    \\
stories\_activities.json  & -        & -      & -     & -       & -     & 35         & 35    \\\hline
total                     & 312	     & 218	  & 580	  & 177	    & 278	& 8,568      &10,133\\
     \hline
\end{tabular}\\
\caption{Descriptive statistics of visual and textual content in the generated Instagram DDP data-set}
\label{tab:descriptive}
\end{table}

The final data-set comprised $11$ Instagram DDPs, containing a total of $723$ .JPEG files (images) on which $1,336$ faces were identified and $304$ usernames and $107$ videos on which $164$ faces were identified and $126$ usernames. In addition, the .json files contain $8,866$ usernames, $904$ first names, $218$ e-mail addresses, $178$ phone numbers and $278$ URLS. See Table \ref{tab:descriptive} for more detailed descriptive statistics regarding the visual content of the generated Instagram DDPs data-set.

\subsection{Approach for textual content}
To evaluate the performance of the de-identification procedure in terms of textual content we consider PII in the form of usernames, first names, e-mail addresses, phone numbers and URLS. 

\begin{figure}[t]
\includegraphics[scale=0.7]{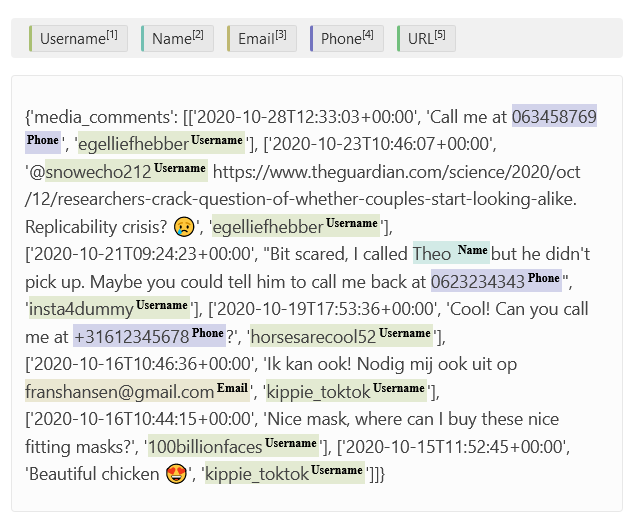}
\caption{An example of how labeling a comments.json file would look like in Label-Studio.}\label{fig:labelstudio}
\end{figure}

The first step of the evaluation procedure is establishing a \textit{ground truth}. Using the $11$ Instagram DDPs, a human rater had to manually label all PII categories per text file, per DDP\footnote{N.B. Establishing the ground truth only has to be done once. The labeling output, together with the 11 Instagram DDPs, are made available for research.}. To make the counting of the labels more efficient and less prone to errors, the labeling process was done in Label-Studio (Figure \ref{fig:labelstudio}). 

\begin{figure}[t]
\includegraphics[scale=0.5]{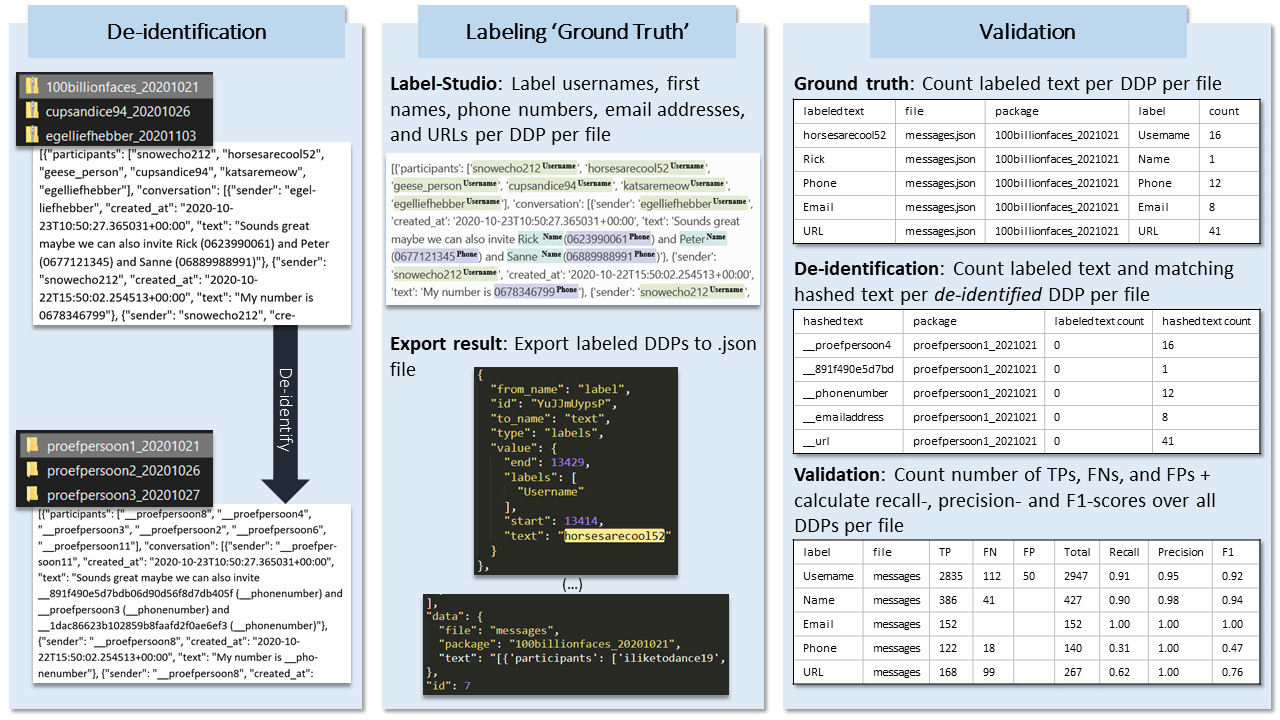}
\caption{The raw DDPs in which all PII categories are labeled (i.e., the ground truth) is compared with the de-identified DDPs. The software counts the number of PII categories (total), correctly hashed PII (TP), falsely hashed information (FP), and unhashed PII (FN). Subsequently, a recall-, precision-, and F1-score can be calculated.}\label{fig:validation}
\end{figure}

Label-Studio returns an output file (result.json) that consists of multiple dictionaries; one per file (e.g., `messages.json'), per package (e.g., `100billionfaces\_20201021'). These dictionaries contain all the labeled text-items (e.g., `horsesarecool52') and corresponding labels (e.g., `Username') present in that specific file (Figure \ref{fig:validation}).

Based on the ground truth, the number of PII categories per text file, per DDP can be determined. Next, using the key files created in the de-identification process, the number of corresponding hashes present in the de-identified DDPs are also calculated per text file, per DDP.

Comparing the PII occurences in the raw DDPs with the PII and corresponding hash occurences, the software can determine the number of times a type of PII was correctly de-identified (True Positive, TP), the number of times a piece of text was incorrectly de-identified (False Positive, FP) and the number of times PII was not de-identified (False Negative, FN). Finally, the recall-, precision-, and F1-score are calculated.

The username is the most important type of PII in DDPs, this holds for Instagram but for DDPs of many other data controllers as well, as usernames are typically unique and can be related to the data subject directly. The software distinguishes between two types of usernames. The researcher can provide a list with usernames of all research participants, and these usernames should be replaced with participant numbers (first type). The second type are all other usernames that appear in the DDPs and those should be replaced by a unique identification code. For both types it holds that they can by correctly de-identified (TP), not be de-identified (FN) or a random piece of text can be replaced by the participant number of the hash (FP). In addition, when a username of a participant is replaced by a wrong participant number or a unique identification code, this is also considered a FN. Researchers intended to use this software can decide for themselves if they want to include a list with participants.

First names should be replaced by a unique identification code (TP). If first names are not replaced they are flagged as falve negatives. In addition, false positives can occur, for example if a hash is applied to a word that is mistaken for a first name, such as the word ``ben" in the Dutch sentence ``Ik ben vandaag jarig." In addition to the list containing the $10.000$ most frequently used Dutch first names that has been used in the EHR de-identification software DEDUCE \citep{menger2018deduce}, we added the first names of the research participants to the list. Furthermore, the software allows you to decide if you want to hash only names that appear in the names list and that start with a capital in the DDP, or if you also want to hash names that do not start with a capital.

\subsection{Approach for visual content}
To annotate visual content, a procedure was carried our by hand, as for each file it had to be determined whether there were one or multiple identifiable faces present and for each detected face whether it was indeed de-identified by the software. To determine whether a face was identifiable, we used a pragmatic definition where we defined a faces as identifiable if at least three out of five facial landmarks were visible (right eye, left eye, nose, right mouth corner and left mouth corner) \cite{zhang2016facial}. This definition will not hold if a person will for example actively try to identify individuals by combining multiple images where a person is partly visible, but it provides a sufficient quality in the sense that in case of a data leak, the person on the images is not directly identified. 

For each piece of visual content it holds that each identified face is considered a single observation which can be either appropriately de-identified (TP) or not (FN). Note that although a video consists of multiple frames in which the possibility arises that a face is identifiable, an instance of one frame showing an identifiable face following our definition results in one FN for this face in the movie. As the determination of whether a face is defined identifiable or not is performed by a human rater and this distinction is sometimes not straightforward, the questionable cases are independently rated by two raters and classification is performed based on consensus. In addition, a set of $100$ .JPEG files and $20$ .MP4 files were independently annotated by two separate annotators. On the .JPEG files, $204$ faces were identified and from these, $193$ were identified by both raters, which equals $94.6\%$. On this subset, a Cohen's $\kappa$ inter-rater reliability was calculated of $1$, so the raters highly agreed on which faces were appropriately de-identified and which not. For the .MP4 files, $49$ faces were identified and from these, $41$ were identified by both raters, which equals $83.7\%$. On this subset, a Cohen's $\kappa$ inter-rater reliability was calculated of $0.62$. The sample of faces was much smaller for .MP4 compared to .JPEG, and it was apparently also a lot more difficult to determine whether a face was appropriately identified when the image was moving compared to when it was a still image. 

In addition, particularly on Instagram, visual content can contain usernames. The software is not able to distinguish between usernames and other types of text, and therefore usernames on visual content can only be detected and de-identified, distinctions between research participants and other usernames are not made. Therefore, appropriately de-identified usernames are counted as true positives (TP) and usernames not de-identified are counted as false negatives (FN). False positives cannot be quantified in the current procedure. 

\subsection{Evaluation criteria}
For each category of PII in each filetype in the set of DDPs regarding textual content, we count the number of TP, FP and FN. For the visual content, we calculate the TP and FN. We use scikit learn to further evaluate the performance of the procedure on the different aspects \citep{pedregosa2011scikit}. First, we calculate the recall, or the sensitivity, as 
\begin{equation}
    Recall = \frac{TP}{TP + FN}.
\end{equation}
Here, we measure the ratio of the correctly de-identified cases to all the cases that were supposed to be de-identified (i.e. ground truth). Each false negative potentially results in not preserving the privacy of a research participant and therefore a high value for the recall is particularly important. The precision is calculated as
\begin{equation}
    Precision = \frac{TP }{TP + FP}.
\end{equation}
Precision shows the ratio of correctly de-identified observations to the total of de-identified observations and a high precision illustrates that the amount of additional information lost due to unnecessary de-identification is limited. Given that DDPs are typically collected to analyze aspects such as the free text or the images, losing a lot of this information by the de-identification process challenges the intended research goal. At last, we calculate the F1 score
\begin{equation}
    F1-score = 2 \times \frac{precision \times recall}{precision + recall},
\end{equation}
which combined the precision and recall and considered both false positives and false negatives. Note that we do not calculate the accuracy as the number of true negatives cannot be determined appropriately in our data-set. 

\section{Results}
\subsection{Initial Results}
In Table \ref{tab:results1}, the results of the application of the software to our Instagram DDP data-set can be found, where we chose for settings including a participant file and capital sensitivity for first names. Regarding the visual content, we can conclude a large proportion of faces on images is appropriately detected and blurred, while on videos this proportion is substantively lower. Apparently, faces are harder to detect by the detection algorithm when the images are moving. 

Regarding textual content, we can conclude that email addresses are appropriately detected and anonymized throughout all files within the DDPs. Regarding names, phone numbers and URLs, we can conclude that a substantial amount of names are not detected by the algorithm throughout the different files. The quality of the anonymization of usernames differs a lot depending on the file. Only in the file `messages.json', false positives are detected. Furthermore, relatively lower recall values are measured for the files `media.json' and `saved.json', although these files have a small number of total observations.

By critically investigating the results found in Table \ref{tab:results1}, and investigating what coding decisions led to the most (negatively) outstanding results, improvements to the code were made. 

\begin{table}
\begin{tabular}{llrrrrrrr}
\hline \hline
Visual \\ \hline
&&            Total & TP    & FN   & FP  & Recall    & Precision & F1  \\ \hline
Faces \\
& .JPEG     & 1,356  & 1,205   & 151  & -  & 0.89      & -         & -   \\
& .MP4      & 199   & 131    & 68   & -  & 0.66      & -         & -   \\ \hline
& Total     & 1,555  & 1,336   & 219  & -  & 0.86      & -         & -   \\
Usernames \\
& .JPEG     & 304   & 302    & 2    & -  & 0.99      & -         & -   \\
& .MP4      & 126   & 125    & 1    & -  & 0.99      & -         & -   \\ \hline
& Total     & 430   & 427    & 3    & -  & 0.99      & -         & -   \\ \hline \hline
Textual \\ \hline
&	file	&	total	&	TP	&	FN	&	FP	&	Recall	&	Precision	&	F1	\\
Email \\
&	comments.json	&	28	&	28	&	0	&	0	&	1	&	1	&	1	\\
&	media.json	    &	28	&	28	&	0	&	0	&	1	&	1	&	1	\\
&	messages.json	&	152	&	152	&	0	&	0	&	1	&	1	&	1	\\
&	profile.json	&	10	&	10	&	0	&	0	&	1	&	1	&	1	\\ \hline
&	total	        &	218	&	218	&	0	&	0	&	1	&	1	&	1	\\ \hline \hline
Name	\\
&	comments.json	&	105	&	61	&	44	&	0	&	0.5619	&	0.9365	&	0.7024	\\
&	media.json	    &	54	&	41	&	13	&	0	&	0.7593	&	1	&	0.8530	\\
&	messages.json	&	427	&	386	&	41	&	0	&	0.9040	&	0.9836	&	0.9374	\\
&	profile.json	&	10	&	6	&	4	&	0	&	0.6	&	1	&	0.75	\\ \hline
&	total	        &	596	&	494	&	102	&	0	&	0.8255	&	0.9798	&	0.8936	\\ \hline \hline
Phone	\\
&	comments.json	&	29	&	26	&	3	&	0	&	0.4828	&	1	&	0.6512	\\
&	media.json	    &	9	&	7	&	2	&	0	&	0.4444	&	1	&	0.6154	\\
&	messages.json	&	139	&	121	&	18	&	0	&	0.3022	&	1	&	0.4641	\\ \hline
&	total	        &	177	&	154	&	23	&	0	&	0.3390	&	1	&	0.5063	\\ \hline \hline
URL	\\
&	comments.json	&	1	&	0	&	1	&	0	&	0	&	0	&	0	\\
&	messages.json	&	267	&	168	&	99	&	0	&	0.6180	&	1	&	0.7639	\\
&	profile.json	&	10	&	10	&	0	&	0	&	1	&	1	&	1	\\ \hline
&	total	        &	278	&	178	&	100	&	0	&	0.6295	&	1	&	0.7726	\\ \hline \hline
Username	\\
&	comments.json	&	261	&	252	&	9	&	0	&	0.9655	&	1	&	0.9813	\\
&	connections.json &	1,222	&	1,190	&	32	&	0	&	0.9722	&	1	&	0.9858	\\
&	likes.json	    &	883	&	823	&	60	&	0	&	0.9320	&	1	&	0.9611	\\
&	media.json	    &	43	&	33	&	10	&	0	&	0.7674	&	0.7907	&	0.7788	\\
&	messages.json	&	2,947	&	2,835	&	112	&	50	&	0.9067	&	0.9500	&	0.9196	\\
&	profile.json	&	10	&	10	&	0	&	0	&	1	&	1	&	1	\\
&	saved.json	&	6	&	4	&	2	&	0	&	0.6667	&	1	&	0.8	\\
&	searches.json	&	314	&	305	&	9	&	0	&	0.9713	&	1	&	0.9855	\\
&	seen\_content.json &	3,144	&	2,619	&	525	&	0	&	0.8330	&	0.9876	&	0.8931 \\
&	shopping.json	&	1	&	1	&	0	&	0	&	1	&	1	&	1	\\
&	stories\_activities.json	&	35	&	34	&	1	&	0	&	0.9714	&	1	&	0.9851	\\ \hline
&	total	&	8,866	&	8,106	&	760	&	50	&	0.89567	&	0.9775	&	0.9324	\\ \hline \hline
\end{tabular}\\
\caption{Results in terms of TP, FP, FN, recall, precision and F1.}
\label{tab:results1}
\end{table}

\subsection{Further improvements}
The first improvement relates to the `profile.json' file. Here, the entire entry that can be found after `name' is now added to the key file and the similar key is used for the DDP username. In this way, the participant can be recognized throughout the complete DDP with either their username of their name. A second improvement has been made after further inspecting the relatively large amount of false positives in the `seen\_content.json' file. Based on this, the list of labels that should be exempted from hashing has been extended. Based on a more thorough inspection of the type of usernames that were not detected by the algorithm, the username format has been adjusted in such a way that usernames are detected as such when they contain at least three characters, the minimum limit in the previous version of the code as six characters. After further inspecting the false positive first names, the names `Van', `Door' and `Can' were removed from the list with the $10,000$ most frequently used first names because they also represent words commonly used in free text, resulting in a lot of FPs. At last, the hash function for usernames became case insensitive, as Instagram does not distinguish between lowercases and uppercases in usernames, while the software initially generated a different hash as an uppercase was used somewhere in the username compared to the username without uppercase. 

The improved script has fewer false negatives regarding names, phone numbers and URLS. Regarding usernames, both the number of false negatives and false positives has decreased substantively. 

\begin{table}
\begin{tabular}{llrrrrrrr}
\hline \hline
	&	file	&	total	&	TP	&	FN	&	FP	&	Recall	&	Precision	&	F1	\\ \hline
DDP\_id \\	
&	messages.json	&	294	&	294	&	0	&	0	&	1	&	1	&	1	\\
&	profile.json	&	18	&	18	&	0	&	0	&	1	&	1	&	1	\\ \hline
&	total	&	312	&	312	&	0	&	0	&	1	&	1	&	1	\\ \hline \hline
E-mail \\
&	comments.json	&	28	&	28	&	0	&	0	&	1	&	1	&	1	\\
&	media.json	&	28	&	28	&	0	&	0	&	1	&	1	&	1	\\
&	messages.json	&	152	&	152	&	0	&	0	&	1	&	1	&	1	\\
&	profile.json	&	10	&	10	&	0	&	0	&	1	&	1	&	1	\\ \hline
&	total	&	218	&	218	&	0	&	0	&	1	&	1	&	1	\\ \hline \hline
Name \\
&	comments.json	&	105	&	98	&	7	&	0	&	0.9333	&	1	&	0.9654	\\
&	media.json	&	54	&	45	&	9	&	0	&	0.8333 &	1	&	0.9042	\\
&	messages.json	&	421	&	385	&	36	&	0	&	0.9145	&	1	&	0.9509	\\ \hline
&	total	&	580	&	528	&	52	&	0	&	0.9103	&	1	&	0.9519	\\ \hline \hline
Phone	\\
&	comments.json	&	29	&	29	&	0	&	0	&	1	&	1	&	1	\\
&	media.json	&	9	&	9	&	0	&	0	&	1	&	1	&	1	\\
&	messages.json	&	139	&	138	&	1	&	24	&	0.9928	&	0.8519	&	0.9169	\\  \hline
&	total	&	177	&	176	&	1	&	24	&	0.9943 &	0.88	&	0.9337	\\ \hline \hline
URL\\
&	comments.json	&	1	&	1	&	0	&	0	&	1	&	1	&	1	\\
&	messages.json	&	267	&	267	&	0	&	0	&	1	&	1	&	1	\\
&	profile.json	&	10	&	10	&	0	&	0	&	1	&	1	&	1	\\ \hline
&	total	&	278	&	278	&	0	&	0	&	1	&	1	&	1	\\ \hline \hline
Username \\
&	comments.json	&	261	&	258	&	3	&	0	&	0.9885	&	1	&	0.9940	\\
&	connections.json	&	1,222	&	1,219	&	3	&	0	&	0.9975	&	1	&	0.9988	\\
&	likes.json	&	883	&	881	&	2	&	0	&	0.9977	&	1	&	0.9989	\\
&	media.json	&	43	&	42	&	1	&	0	&	0.9767	&	1	&	0.9881	\\
&	messages.json	&	2,659	&	2,658	&	1	&	2	&	0.9846	&	0.9868	&	0.9847	\\
&	profile.json	&	1	&	1	&	0	&	1	&	0	&	0	&	0	\\
&	saved.json	&	6	&	6	&	0	&	0	&	1	&	1	&	1	\\
&	searches.json	&	314	&	313	&	1	&	0	&	0.9968	&	1	&	0.9984	\\
&	seen\_content.json	&	3,143	&	3,137	&	6	&	0	&	0.9981	&	1	&	0.9990	\\
&	shopping.json	&	1	&	1	&	0	&	0	&	1	&	1	&	1	\\
&	stories\_activities.json	&	35	&	35	&	0	&	0	&	1	&	1	&	1	\\ \hline
&	total	&	8,568	&	8,551	&	17	&	3	&	0.9932	&	0.9985	&	0.9952	\\ \hline \hline
\end{tabular}\\
\caption{Results in terms of TP, FP, FN, recall, precision and F1 after improvements to the script have been made.}
\label{tab:results2}
\end{table}

\section{Conclusions and future work}
Data Download Packages (DDPs) contain all data collected by public and private entities during the course of citizens' digital life. Although they form a treasure trove for social scientists, they contain data that can be deeply private. To protect the privacy of research participants while they let their DDPs be used for scientific research, we developed de-identification software that is able to anonymize and pseudonymize data that follow typical DDP structures.

We evaluated the performance of the de-identification software on a set of Instagram DDPs. From this application we could conclude that the software is particularly well suited to anonymize and/or pseudonymize usernames, e-mail addresses and phone-numbers from structured and unstructured text files. In addition, it was able to appropriately anonymize faces on .jpg files. Appropriate anonymization and/or pseudonymization of first names appeared more challenging, particularly because some first names can also appear as words in open text and vice versa. However, when applying the software researchers can decide if their focus is on precision or on recall and take measures to accomodate this. Furthermore, anonymizing faces on .mp4 files appeared more challenging, typically because in moving images sometimes different parts of faces can be visible at different moments, together providing sufficient information to identify a face, and because Instagram provides in so-called 'filters', which also make it more difficult for the software to detect a face for de-identification. 

The aim of the software was to remove identifiers from DDPs in such a way that research participants cannot be identified when the data is manually investigated or in the undesired situation that someone gains unauthorized access to the data. Appropriate safety measures to prevent this remain required, but based on the results from the validation we do believe that the intended goal of this software is met. 

If researchers intend to use this software for their own research projects, a number of issues should be taken into account. A first issue is that the current script has been primarily be developed to de-identify the Instagram DDPs. However, the software has been written in such a way that with small adjustments it could be applied to DDPs from other data controllers. In future work, we could provide some of these adjustments for specific other data controllers to illustrate how this works in practice, but we also encourage other researchers and software developers to develop such adjustments and share this with the community. A second issue is that, besides adjustments to DDPs from different data controllers, we can also imagine that different researchers might have different research intentions with the collected data and that based on this adjustments to the software might be desired. For example a sociologist with interest in what types of accounts are followed and liked by the research participant might not want to pseudonymize all usernames present in the DDP, but instead only the usernames of the participants for example. A third issue to consider is that if a higher level of security is desired, adjustments can also be made in a quite straightforward manner. For example, it can be chosen not to save the key file or to use hashing and blurring algorithms with higher safety standards. 

An important issue to note further is that because of the fact that faces on images are blurred when this software is used, it is no longer possible to for example apply emotion detection algorithms to the faces on the images in the DDPs under investigation. If emotion detection of faces is a goal of the researcher, it can be considered to replace the blurring part of the software with a procedure that replaces the face with a deepfake of the face \citep{korshunov2018deepfakes}. With such an algorithm, it remains possible to detect the emotions on faces, while protecting the privacy of the participants. However, this will inevitably also introduce some noise. 

Another remark regarding the blurring of visual content is that this part of the software could be further developed to be more refined so that it can distinguish between usernames and regular text and that it only blurs the usernames. In addition, it can be further refined in such a way that text written for example at a 45$^{\circ}$ or 90$^{\circ}$ is evaluated in a single sequence as well. currently, angled text is typically evaluated in small separate pieces. A last point of attention is that sound in .mp4 files is currently removed. This might be a good thing as it thereby also removes possibly identifying sounds but it might be disadvantageous for certain purposes. Although the use of digital trace data for scientific purposes, and appropriate de-identification of digital trace data are fields that are still at their infancy, our developed software enormously contributes to privacy preserving analysis of digital trace data collected with DDPs.   

\bibliography{bibliography}

\begin{thebibliography}{}

\bibitem [\protect \citeauthoryear {%
Azure%
}{%
Azure%
}{%
{\protect \APACyear {2021}}%
}]{%
Microsoft2021Azure}
\APACinsertmetastar {%
Microsoft2021Azure}%
\begin{APACrefauthors}%
Azure, M.%
\end{APACrefauthors}%
\unskip\
\newblock
\APACrefYearMonthDay{2021}{Jan}{}.
\newblock
\APACrefbtitle {Microsoft Azure cognitive services.} {Microsoft azure cognitive
  services.}
\newblock
\begin{APACrefURL}
  \url{https://azure.microsoft.com/nl-nl/services/cognitive-services/face/}
  \end{APACrefURL}
\PrintBackRefs{\CurrentBib}

\bibitem [\protect \citeauthoryear {%
Boeschoten%
, Ausloos%
, Moeller%
, Araujo%
\BCBL {}\ \BBA {} Oberski%
}{%
Boeschoten%
\ \protect \BOthers {.}}{%
{\protect \APACyear {2020}}%
}]{%
boeschoten2020digital}
\APACinsertmetastar {%
boeschoten2020digital}%
\begin{APACrefauthors}%
Boeschoten, L.%
, Ausloos, J.%
, Moeller, J.%
, Araujo, T.%
\BCBL {}\ \BBA {} Oberski, D\BPBI L.%
\end{APACrefauthors}%
\unskip\
\newblock
\APACrefYearMonthDay{2020}{}{}.
\newblock
{\BBOQ}\APACrefatitle {Digital trace data collection through data donation}
  {Digital trace data collection through data donation}.{\BBCQ}
\newblock
\APACjournalVolNumPages{arXiv preprint arXiv:2011.09851}{}{}{}.
\PrintBackRefs{\CurrentBib}

\bibitem [\protect \citeauthoryear {%
Bradski%
}{%
Bradski%
}{%
{\protect \APACyear {2000}}%
}]{%
opencv2000}
\APACinsertmetastar {%
opencv2000}%
\begin{APACrefauthors}%
Bradski, G.%
\end{APACrefauthors}%
\unskip\
\newblock
\APACrefYearMonthDay{2000}{}{}.
\newblock
{\BBOQ}\APACrefatitle {{The OpenCV Library}} {{The OpenCV Library}}.{\BBCQ}
\newblock
\APACjournalVolNumPages{Dr. Dobb's Journal of Software Tools}{}{}{}.
\newblock
\begin{APACrefURL} \url{https://opencv.org/} \end{APACrefURL}
\PrintBackRefs{\CurrentBib}

\bibitem [\protect \citeauthoryear {%
Coppersmith%
, Mitchell%
, Harman%
, Dredze%
\BCBL {}\ \BBA {} Leary%
}{%
Coppersmith%
\ \protect \BOthers {.}}{%
{\protect \APACyear {2017}}%
}]{%
Coppersmith2017Twitter}
\APACinsertmetastar {%
Coppersmith2017Twitter}%
\begin{APACrefauthors}%
Coppersmith, G.%
, Mitchell, M.%
, Harman, C.%
, Dredze, M.%
\BCBL {}\ \BBA {} Leary, R.%
\end{APACrefauthors}%
\unskip\
\newblock
\APACrefYearMonthDay{2017}{Feb}{}.
\newblock
\APACrefbtitle {Deidentify Twitter.} {Deidentify twitter.}
\newblock
\begin{APACrefURL} \url{https://github.com/qntfy/deidentify_twitter}
  \end{APACrefURL}
\PrintBackRefs{\CurrentBib}

\bibitem [\protect \citeauthoryear {%
El~Emam%
\ \BBA {} Dankar%
}{%
El~Emam%
\ \BBA {} Dankar%
}{%
{\protect \APACyear {2008}}%
}]{%
el2008protecting}
\APACinsertmetastar {%
el2008protecting}%
\begin{APACrefauthors}%
El~Emam, K.%
\BCBT {}\ \BBA {} Dankar, F\BPBI K.%
\end{APACrefauthors}%
\unskip\
\newblock
\APACrefYearMonthDay{2008}{}{}.
\newblock
{\BBOQ}\APACrefatitle {Protecting privacy using k-anonymity} {Protecting
  privacy using k-anonymity}.{\BBCQ}
\newblock
\APACjournalVolNumPages{Journal of the American Medical Informatics
  Association}{15}{5}{627--637}.
\newblock
\begin{APACrefURL} \url{https://doi.org/10.1197/jamia.M2716} \end{APACrefURL}
\PrintBackRefs{\CurrentBib}

\bibitem [\protect \citeauthoryear {%
Esler%
}{%
Esler%
}{%
{\protect \APACyear {2019}}%
}]{%
esler2019}
\APACinsertmetastar {%
esler2019}%
\begin{APACrefauthors}%
Esler, T.%
\end{APACrefauthors}%
\unskip\
\newblock
\APACrefYearMonthDay{2019}{Dec}{}.
\newblock
\APACrefbtitle {facenet pytorch.} {facenet pytorch.}
\newblock
\begin{APACrefURL} \url{https://www.kaggle.com/timesler/facenet-pytorch}
  \end{APACrefURL}
\newblock
\begin{APACrefDOI} \doi{10.34740/KAGGLE/DSV/845275} \end{APACrefDOI}
\PrintBackRefs{\CurrentBib}

\bibitem [\protect \citeauthoryear {%
Heider%
, Obeid%
\BCBL {}\ \BBA {} Meystre%
}{%
Heider%
\ \protect \BOthers {.}}{%
{\protect \APACyear {2020}}%
}]{%
heider2020comparative}
\APACinsertmetastar {%
heider2020comparative}%
\begin{APACrefauthors}%
Heider, P\BPBI M.%
, Obeid, J\BPBI S.%
\BCBL {}\ \BBA {} Meystre, S\BPBI M.%
\end{APACrefauthors}%
\unskip\
\newblock
\APACrefYearMonthDay{2020}{}{}.
\newblock
{\BBOQ}\APACrefatitle {A Comparative Analysis of Speed and Accuracy for Three
  Off-the-Shelf De-Identification Tools} {A comparative analysis of speed and
  accuracy for three off-the-shelf de-identification tools}.{\BBCQ}
\newblock
\APACjournalVolNumPages{AMIA Summits on Translational Science
  Proceedings}{2020}{}{241}.
\newblock
\begin{APACrefDOI} \doi{PMCID: PMC7233098} \end{APACrefDOI}
\PrintBackRefs{\CurrentBib}

\bibitem [\protect \citeauthoryear {%
Jungherr%
}{%
Jungherr%
}{%
{\protect \APACyear {2015}}%
}]{%
jungherrAnalyzingPoliticalCommunication2015}
\APACinsertmetastar {%
jungherrAnalyzingPoliticalCommunication2015}%
\begin{APACrefauthors}%
Jungherr, A.%
\end{APACrefauthors}%
\unskip\
\newblock
\APACrefYear{2015}.
\newblock
\APACrefbtitle {Analyzing {Political} {Communication} with {Digital} {Trace}
  {Data}: {The} {Role} of {Twitter} {Messages} in {Social} {Science}
  {Research}} {Analyzing {Political} {Communication} with {Digital} {Trace}
  {Data}: {The} {Role} of {Twitter} {Messages} in {Social} {Science}
  {Research}}.
\newblock
\APACaddressPublisher{}{Springer International Publishing}.
\newblock
\begin{APACrefURL}
  [{2020-08-04}]\url{https://www.springer.com/gp/book/9783319203188}
  \end{APACrefURL}
\newblock
\begin{APACrefDOI} \doi{10.1007/978-3-319-20319-5} \end{APACrefDOI}
\PrintBackRefs{\CurrentBib}

\bibitem [\protect \citeauthoryear {%
King%
}{%
King%
}{%
{\protect \APACyear {2011}}%
}]{%
king2011ensuring}
\APACinsertmetastar {%
king2011ensuring}%
\begin{APACrefauthors}%
King, G.%
\end{APACrefauthors}%
\unskip\
\newblock
\APACrefYearMonthDay{2011}{}{}.
\newblock
{\BBOQ}\APACrefatitle {Ensuring the data-rich future of the social sciences}
  {Ensuring the data-rich future of the social sciences}.{\BBCQ}
\newblock
\APACjournalVolNumPages{science}{331}{6018}{719--721}.
\newblock
\begin{APACrefDOI} \doi{10.1126/science.1197872} \end{APACrefDOI}
\PrintBackRefs{\CurrentBib}

\bibitem [\protect \citeauthoryear {%
Korshunov%
\ \BBA {} Marcel%
}{%
Korshunov%
\ \BBA {} Marcel%
}{%
{\protect \APACyear {2018}}%
}]{%
korshunov2018deepfakes}
\APACinsertmetastar {%
korshunov2018deepfakes}%
\begin{APACrefauthors}%
Korshunov, P.%
\BCBT {}\ \BBA {} Marcel, S.%
\end{APACrefauthors}%
\unskip\
\newblock
\APACrefYearMonthDay{2018}{}{}.
\newblock
{\BBOQ}\APACrefatitle {Deepfakes: a new threat to face recognition? assessment
  and detection} {Deepfakes: a new threat to face recognition? assessment and
  detection}.{\BBCQ}
\newblock
\APACjournalVolNumPages{arXiv preprint arXiv:1812.08685}{}{}{}.
\PrintBackRefs{\CurrentBib}

\bibitem [\protect \citeauthoryear {%
Kosinski%
, Stillwell%
\BCBL {}\ \BBA {} Graepel%
}{%
Kosinski%
\ \protect \BOthers {.}}{%
{\protect \APACyear {2013}}%
}]{%
kosinskiPrivateTraitsAttributes2013}
\APACinsertmetastar {%
kosinskiPrivateTraitsAttributes2013}%
\begin{APACrefauthors}%
Kosinski, M.%
, Stillwell, D.%
\BCBL {}\ \BBA {} Graepel, T.%
\end{APACrefauthors}%
\unskip\
\newblock
\APACrefYearMonthDay{2013}{{\APACmonth{04}}}{}.
\newblock
{\BBOQ}\APACrefatitle {Private traits and attributes are predictable from
  digital records of human behavior} {Private traits and attributes are
  predictable from digital records of human behavior}.{\BBCQ}
\newblock
\APACjournalVolNumPages{Proceedings of the National Academy of
  Sciences}{110}{15}{5802--5805}.
\newblock
\begin{APACrefURL}
  [{2020-08-04}]\url{http://www.pnas.org/cgi/doi/10.1073/pnas.1218772110}
  \end{APACrefURL}
\newblock
\begin{APACrefDOI} \doi{10.1073/pnas.1218772110} \end{APACrefDOI}
\PrintBackRefs{\CurrentBib}

\bibitem [\protect \citeauthoryear {%
Kushida%
\ \protect \BOthers {.}}{%
Kushida%
\ \protect \BOthers {.}}{%
{\protect \APACyear {2012}}%
}]{%
kushida2012strategies}
\APACinsertmetastar {%
kushida2012strategies}%
\begin{APACrefauthors}%
Kushida, C\BPBI A.%
, Nichols, D\BPBI A.%
, Jadrnicek, R.%
, Miller, R.%
, Walsh, J\BPBI K.%
\BCBL {}\ \BBA {} Griffin, K.%
\end{APACrefauthors}%
\unskip\
\newblock
\APACrefYearMonthDay{2012}{}{}.
\newblock
{\BBOQ}\APACrefatitle {Strategies for de-identification and anonymization of
  electronic health record data for use in multicenter research studies}
  {Strategies for de-identification and anonymization of electronic health
  record data for use in multicenter research studies}.{\BBCQ}
\newblock
\APACjournalVolNumPages{Medical care}{50}{Suppl}{S82}.
\newblock
\begin{APACrefDOI} \doi{10.1097/MLR.0b013e3182585355} \end{APACrefDOI}
\PrintBackRefs{\CurrentBib}

\bibitem [\protect \citeauthoryear {%
Liu%
, Perez-Concha%
, Nguyen%
, Bennett%
\BCBL {}\ \BBA {} Jorm%
}{%
Liu%
\ \protect \BOthers {.}}{%
{\protect \APACyear {2020}}%
}]{%
liu2020identifying}
\APACinsertmetastar {%
liu2020identifying}%
\begin{APACrefauthors}%
Liu, L.%
, Perez-Concha, O.%
, Nguyen, A.%
, Bennett, V.%
\BCBL {}\ \BBA {} Jorm, L.%
\end{APACrefauthors}%
\unskip\
\newblock
\APACrefYearMonthDay{2020}{}{}.
\newblock
{\BBOQ}\APACrefatitle {De-identifying Hospital Discharge Summaries: An
  End-to-End Framework using Ensemble of De-Identifiers} {De-identifying
  hospital discharge summaries: An end-to-end framework using ensemble of
  de-identifiers}.{\BBCQ}
\newblock
\APACjournalVolNumPages{arXiv preprint arXiv:2101.00146}{}{}{}.
\PrintBackRefs{\CurrentBib}

\bibitem [\protect \citeauthoryear {%
Menger%
, Scheepers%
, van Wijk%
\BCBL {}\ \BBA {} Spruit%
}{%
Menger%
\ \protect \BOthers {.}}{%
{\protect \APACyear {2018}}%
}]{%
menger2018deduce}
\APACinsertmetastar {%
menger2018deduce}%
\begin{APACrefauthors}%
Menger, V.%
, Scheepers, F.%
, van Wijk, L\BPBI M.%
\BCBL {}\ \BBA {} Spruit, M.%
\end{APACrefauthors}%
\unskip\
\newblock
\APACrefYearMonthDay{2018}{}{}.
\newblock
{\BBOQ}\APACrefatitle {DEDUCE: A pattern matching method for automatic
  de-identification of Dutch medical text} {Deduce: A pattern matching method
  for automatic de-identification of dutch medical text}.{\BBCQ}
\newblock
\APACjournalVolNumPages{Telematics and Informatics}{35}{4}{727--736}.
\newblock
\begin{APACrefURL} \url{https://doi.org/10.1016/j.tele.2017.08.002}
  \end{APACrefURL}
\PrintBackRefs{\CurrentBib}

\bibitem [\protect \citeauthoryear {%
Pedregosa%
\ \protect \BOthers {.}}{%
Pedregosa%
\ \protect \BOthers {.}}{%
{\protect \APACyear {2011}}%
}]{%
pedregosa2011scikit}
\APACinsertmetastar {%
pedregosa2011scikit}%
\begin{APACrefauthors}%
Pedregosa, F.%
, Varoquaux, G.%
, Gramfort, A.%
, Michel, V.%
, Thirion, B.%
, Grisel, O.%
\BDBL {}others%
\end{APACrefauthors}%
\unskip\
\newblock
\APACrefYearMonthDay{2011}{}{}.
\newblock
{\BBOQ}\APACrefatitle {Scikit-learn: Machine learning in Python} {Scikit-learn:
  Machine learning in python}.{\BBCQ}
\newblock
\APACjournalVolNumPages{the Journal of machine Learning
  research}{12}{}{2825--2830}.
\newblock
\begin{APACrefURL}
  \url{https://www.jmlr.org/papers/volume12/pedregosa11a/pedregosa11a.pdf?source=post_page---------------------------}
  \end{APACrefURL}
\PrintBackRefs{\CurrentBib}

\bibitem [\protect \citeauthoryear {%
Regulation%
}{%
Regulation%
}{%
{\protect \APACyear {2016}}%
}]{%
regulation2016regulation}
\APACinsertmetastar {%
regulation2016regulation}%
\begin{APACrefauthors}%
Regulation, G\BPBI D\BPBI P.%
\end{APACrefauthors}%
\unskip\
\newblock
\APACrefYearMonthDay{2016}{}{}.
\newblock
{\BBOQ}\APACrefatitle {Regulation (EU) 2016/679 of the European Parliament and
  of the Council of 27 April 2016 on the protection of natural persons with
  regard to the processing of personal data and on the free movement of such
  data, and repealing Directive 95/46} {Regulation (eu) 2016/679 of the
  european parliament and of the council of 27 april 2016 on the protection of
  natural persons with regard to the processing of personal data and on the
  free movement of such data, and repealing directive 95/46}.{\BBCQ}
\newblock
\APACjournalVolNumPages{Official Journal of the European Union
  (OJ)}{59}{1-88}{294}.
\newblock
\begin{APACrefURL}
  \url{https://eur-lex.europa.eu/legal-content/EN/TXT/PDF/?uri=OJ:L:2016:119:FULL&from=EN}
  \end{APACrefURL}
\PrintBackRefs{\CurrentBib}

\bibitem [\protect \citeauthoryear {%
Schoen%
\ \protect \BOthers {.}}{%
Schoen%
\ \protect \BOthers {.}}{%
{\protect \APACyear {2013}}%
}]{%
schoenPowerPredictionSocial2013}
\APACinsertmetastar {%
schoenPowerPredictionSocial2013}%
\begin{APACrefauthors}%
Schoen, H.%
, Gayo-Avello, D.%
, Takis~Metaxas, P.%
, Mustafaraj, E.%
, Strohmaier, M.%
\BCBL {}\ \BBA {} Gloor, P.%
\end{APACrefauthors}%
\unskip\
\newblock
\APACrefYearMonthDay{2013}{{\APACmonth{10}}}{}.
\newblock
{\BBOQ}\APACrefatitle {The power of prediction with social media} {The power of
  prediction with social media}.{\BBCQ}
\newblock
\APACjournalVolNumPages{Internet Research}{23}{5}{528--543}.
\newblock
\begin{APACrefURL}
  [{2020-08-12}]\url{https://www.emerald.com/insight/content/doi/10.1108/IntR-06-2013-0115/full/html}
  \end{APACrefURL}
\newblock
\begin{APACrefDOI} \doi{10.1108/IntR-06-2013-0115} \end{APACrefDOI}
\PrintBackRefs{\CurrentBib}

\bibitem [\protect \citeauthoryear {%
Simon%
}{%
Simon%
}{%
{\protect \APACyear {2018}}%
}]{%
simon2018amazon}
\APACinsertmetastar {%
simon2018amazon}%
\begin{APACrefauthors}%
Simon, J.%
\end{APACrefauthors}%
\unskip\
\newblock
\APACrefYearMonthDay{2018}{}{}.
\newblock
{\BBOQ}\APACrefatitle {Amazon Comprehend Medical--Natural Language Processing
  24 for Healthcare Customers} {Amazon comprehend medical--natural language
  processing 24 for healthcare customers}.{\BBCQ}
\newblock
\APACjournalVolNumPages{Retrieved April}{18}{}{2019}.
\newblock
\begin{APACrefURL} \url{https://aws.amazon.com/comprehend/medical/}
  \end{APACrefURL}
\PrintBackRefs{\CurrentBib}

\bibitem [\protect \citeauthoryear {%
Sweeney%
}{%
Sweeney%
}{%
{\protect \APACyear {2002}}%
}]{%
sweeney2002k}
\APACinsertmetastar {%
sweeney2002k}%
\begin{APACrefauthors}%
Sweeney, L.%
\end{APACrefauthors}%
\unskip\
\newblock
\APACrefYearMonthDay{2002}{}{}.
\newblock
{\BBOQ}\APACrefatitle {k-anonymity: A model for protecting privacy}
  {k-anonymity: A model for protecting privacy}.{\BBCQ}
\newblock
\APACjournalVolNumPages{International Journal of Uncertainty, Fuzziness and
  Knowledge-Based Systems}{10}{05}{557--570}.
\PrintBackRefs{\CurrentBib}

\bibitem [\protect \citeauthoryear {%
van~der Sloot%
}{%
van~der Sloot%
}{%
{\protect \APACyear {2020}}%
}]{%
van2020general}
\APACinsertmetastar {%
van2020general}%
\begin{APACrefauthors}%
van~der Sloot, B.%
\end{APACrefauthors}%
\unskip\
\newblock
\APACrefYear{2020}.
\newblock
\APACrefbtitle {The General Data Protection Regulation in Plain Language} {The
  general data protection regulation in plain language}.
\newblock
\APACaddressPublisher{}{Amsterdam University Press}.
\PrintBackRefs{\CurrentBib}

\bibitem [\protect \citeauthoryear {%
van Kemenade%
\ \protect \BOthers {.}}{%
van Kemenade%
\ \protect \BOthers {.}}{%
{\protect \APACyear {2020}}%
}]{%
vankemenade2020}
\APACinsertmetastar {%
vankemenade2020}%
\begin{APACrefauthors}%
van Kemenade, H.%
, wiredfool%
, Murray, A.%
, Clark, A.%
, Karpinsky, A.%
, nulano%
\BDBL {}Kurczewski, M.%
\end{APACrefauthors}%
\unskip\
\newblock
\APACrefYearMonthDay{2020}{{\APACmonth{10}}}{}.
\newblock
\APACrefbtitle {python-pillow/Pillow 8.0.0.} {python-pillow/pillow 8.0.0.}
\newblock
\APACaddressPublisher{}{Zenodo}.
\newblock
\begin{APACrefURL} \url{https://doi.org/10.5281/zenodo.4088798}
  \end{APACrefURL}
\newblock
\begin{APACrefDOI} \doi{10.5281/zenodo.4088798} \end{APACrefDOI}
\PrintBackRefs{\CurrentBib}

\bibitem [\protect \citeauthoryear {%
van Toledo%
, van Dijk%
\BCBL {}\ \BBA {} Spruit%
}{%
van Toledo%
\ \protect \BOthers {.}}{%
{\protect \APACyear {{\protect \bibnodate {}}}}%
}]{%
vanevaluating}
\APACinsertmetastar {%
vanevaluating}%
\begin{APACrefauthors}%
van Toledo, C.%
, van Dijk, F.%
\BCBL {}\ \BBA {} Spruit, M.%
\end{APACrefauthors}%
\unskip\
\newblock
\APACrefYearMonthDay{{\protect \bibnodate {}}}{}{}.
\newblock
{\BBOQ}\APACrefatitle {Evaluating Dutch named entity recognition and
  de-identification methods in the human resource domain} {Evaluating dutch
  named entity recognition and de-identification methods in the human resource
  domain}.{\BBCQ}
\newblock

\newblock
\begin{APACrefDOI} \doi{10.5121/csit.2020.101520} \end{APACrefDOI}
\PrintBackRefs{\CurrentBib}

\bibitem [\protect \citeauthoryear {%
Yadong%
}{%
Yadong%
}{%
{\protect \APACyear {2018}}%
}]{%
Yadong2018}
\APACinsertmetastar {%
Yadong2018}%
\begin{APACrefauthors}%
Yadong, O.%
\end{APACrefauthors}%
\unskip\
\newblock
\APACrefYearMonthDay{2018}{Nov}{}.
\newblock
\APACrefbtitle {frozen east text detection.} {frozen east text detection.}
\newblock
\begin{APACrefURL} \url{https://github.com/oyyd/frozen_east_text_detection.pb}
  \end{APACrefURL}
\PrintBackRefs{\CurrentBib}

\bibitem [\protect \citeauthoryear {%
Zhang%
, Zhang%
, Li%
\BCBL {}\ \BBA {} Qiao%
}{%
Zhang%
\ \protect \BOthers {.}}{%
{\protect \APACyear {2016}}%
}]{%
zhang2016facial}
\APACinsertmetastar {%
zhang2016facial}%
\begin{APACrefauthors}%
Zhang, K.%
, Zhang, Z.%
, Li, Z.%
\BCBL {}\ \BBA {} Qiao, Y.%
\end{APACrefauthors}%
\unskip\
\newblock
\APACrefYearMonthDay{2016}{}{}.
\newblock
{\BBOQ}\APACrefatitle {Joint Face Detection and Alignment using Multi-task
  Cascaded Convolutional Networks} {Joint face detection and alignment using
  multi-task cascaded convolutional networks}.{\BBCQ}
\newblock
\APACjournalVolNumPages{IEEE Signal Processing Letters}{23}{10}{1499--1503}.
\newblock
\begin{APACrefDOI} \doi{10.1109/LSP.2016.2603342} \end{APACrefDOI}
\PrintBackRefs{\CurrentBib}

\bibitem [\protect \citeauthoryear {%
Zhong%
, Huang%
\BCBL {}\ \BBA {} Liu%
}{%
Zhong%
\ \protect \BOthers {.}}{%
{\protect \APACyear {2021}}%
}]{%
zhong2021mental}
\APACinsertmetastar {%
zhong2021mental}%
\begin{APACrefauthors}%
Zhong, B.%
, Huang, Y.%
\BCBL {}\ \BBA {} Liu, Q.%
\end{APACrefauthors}%
\unskip\
\newblock
\APACrefYearMonthDay{2021}{}{}.
\newblock
{\BBOQ}\APACrefatitle {Mental health toll from the coronavirus: Social media
  usage reveals Wuhan residents’ depression and secondary trauma in the
  COVID-19 outbreak} {Mental health toll from the coronavirus: Social media
  usage reveals wuhan residents’ depression and secondary trauma in the
  covid-19 outbreak}.{\BBCQ}
\newblock
\APACjournalVolNumPages{Computers in human behavior}{114}{}{106524}.
\newblock
\begin{APACrefURL} \url{https://doi.org/10.1016/j.chb.2020.106524}
  \end{APACrefURL}
\PrintBackRefs{\CurrentBib}

\bibitem [\protect \citeauthoryear {%
Zhou%
\ \protect \BOthers {.}}{%
Zhou%
\ \protect \BOthers {.}}{%
{\protect \APACyear {2017}}%
}]{%
zhou2017east}
\APACinsertmetastar {%
zhou2017east}%
\begin{APACrefauthors}%
Zhou, X.%
, Yao, C.%
, Wen, H.%
, Wang, Y.%
, Zhou, S.%
, He, W.%
\BCBL {}\ \BBA {} Liang, J.%
\end{APACrefauthors}%
\unskip\
\newblock
\APACrefYearMonthDay{2017}{}{}.
\newblock
{\BBOQ}\APACrefatitle {EAST: An Efficient and Accurate Scene Text Detector}
  {East: An efficient and accurate scene text detector}.{\BBCQ}
\newblock
\BIn{} \APACrefbtitle {2017 IEEE Conference on Computer Vision and Pattern
  Recognition (CVPR)} {2017 ieee conference on computer vision and pattern
  recognition (cvpr)}\ (\BPG~2642-2651).
\newblock
\begin{APACrefDOI} \doi{10.1109/CVPR.2017.283} \end{APACrefDOI}
\PrintBackRefs{\CurrentBib}

\end{thebibliography}

\end{document}